\documentclass{cernrep}
\begin{document}
\title{Dielectric Laser Acceleration}
\author{N. Sch\"onenberger and P. Hommelhoff}
\institute{Department of Physics, Friedrich-Alexander-University Erlangen-Nuremberg}

\sloppy
\pagestyle{empty}

\begin{abstract}
Dielectric Laser Accelerators (DLAs) use the nearfields created when a laser pulse impinges on a dielectric structure to accelerate charged particles. We provide an overview of the theory of operation of photon driven accelerators, from photons interacting with charged particles in a vacuum, to the advantages gained by introducing dielectric structures, with a discussion of  their advantages and limitations. Furthermore we show the state of the art of the current development of dielectric laser accelerators, including acceleration, focusing, deflection, beam position monitoring, and advanced topics from the generation of microbunches to the adaptation of alternating phase focusing, allowing for the next to lossless transport of the charged particles over long distances.
\end{abstract}


\maketitle

\section{Introduction}

The idea of accelerators using dielectrics or stimulated emission to generate the necessary electromagnetic fields is not new. First proposals for such machines appeared in the 1970s. These consisted of maser pumped\cite{Shimoda1962} accelerators --- first demonstrated in 1987 using metallic gratings and mm wave radiation ~\cite{Mizuno1987}, laser pumped metallic gratings ~\cite{Takeda1968} and finally laser pumped dielectric ~\cite{lohmann62,lohmann63} accelerators. The first successful demonstrations of what we now call dielectric laser accelerators happened in 2013, independently at relativistic electron velocities ~\cite{Peralta2013} and subrelativistic speeds ~\cite{Breuer}. Several advancements were necessary for a successful demonstration. Among others, the development of ultrashort laser pulses and their amplification --- honoured with the 2018 Nobel prize in physics for the invention of chirped pulse amplification --- to high peak intensities and the development of semiconductor technology that pressed forward the ability of manufacturing tiny devices on a scale of the wavelength used to drive these accelerators. 

Although the use of metal gratings was demonstrated early on, maser or laser driven accelerators suffer the same drawbacks of a rather low damage threshold when using metallic materials as conventional radio frequency (RF) accelerators. Although damage mechanisms can vary from thermal to peak field driven effects metals, due to the availability of excitable electrons, the light can couple to the structure and cause damage. In dielectrics the bandgap provides transparency allowing refractive media to be used instead of metallic ones with much greater peak fields before damage is observed. This facilitates the generation of higher acceleration gradients compared to RF accelerators. \\
Due to the available laser sources that can provide short pulses and therefore sufficiently high peak fields and other considerations such as integrateability and wall plug power efficiency, the prefered wavelengths are usually below \Unit{2000}{nm}. Therefore the structure features are roughly \Unit{1000}{nm} and below. This miniaturization opens the door for applications other accelerator concepts are not capable of supplying \eg medical irradiation in the body. This holds while DLAs currently require roughly an order of magnitude less power per meter of accelerator compared to Laser Plasma Wakefield Accelerators (LPWA). This has still great potential for improvement since most of the laser power is unused and can be recycled improving the DLAs efficiency in future devices. 

\section{Theory}
\subsection{Electromagnetic waves in vacuum}
Starting from Maxwell's equations, the feasibility of acceleration in vacuum, near dielectric boundaries and finally near microstructured dielectrics is explored, highlighting their respective advantages and limitations. In the following section we derive the wave nature of electro magnetic waves. Maxwell's equations
\begin{equation}
\nabla \cdot \vec{E} = \frac{\rho}{\epsilon_0} \label{eq:gauss}
\end{equation}
\begin{equation}
\nabla \cdot \vec{B} = 0
\end{equation}
\begin{equation}
\nabla \times \vec{E} = -\frac{\partial \vec{B}}{\partial t} \label{eq:faraday}
\end{equation}
\begin{equation}
\nabla \times \vec{B} = \mu_0 \epsilon_0 \frac{\partial \vec{E}}{\partial t} + \mu_0 \vec{J} \label{eq:ampere}
\end{equation}
can be simplified with $\rho = 0$ and $\vec{J} = 0$ since all considerations are done in vacuum and therefore there are neither charges nor currents. Taking the curl of Faraday's law of induction Eq.~(\ref{eq:faraday}) changing the right hand side order of the differentiation yields
\begin{equation}
\nabla \times \left( \nabla \times \vec{E} \right) = - \frac{\partial}{\partial t} \left( \nabla \times \vec{B} \right) .
\end{equation}
Substituting Ampere's law Eq.~(\ref{eq:ampere}) and assuming that the vacuum permittivity $\epsilon_0$ and vacuum permeability $\mu_0$ are not time varying we arrive at
\begin{equation}
\nabla \times \left( \nabla \times \vec{E} \right) = - \mu_0 \epsilon_0 \frac{\partial^2\vec{E}}{\partial t^2} .
\end{equation}
Using the identity $\nabla \times \left( \nabla \times \vec{A} \right) = \nabla \left( \nabla \cdot \vec{A} \right) - \nabla^2 \vec{A}$ and Gauss's law Eq.~(\ref{eq:gauss}) yields the wave equation of the electric field 
\begin{equation}
\nabla^2 \vec{E} = \mu_0 \epsilon_0 \frac{\partial^2 \vec{E}}{\partial t^2} .
\end{equation}
Similar steps can be taken starting with Ampere's law to arrive at the analogous wave equation for the magnetic field
\begin{equation}
\nabla^2 \vec{B} = \mu_0 \epsilon_0 \frac{\partial^2 \vec{B}}{\partial t^2} .
\end{equation} 
The easiest solution for these wave equations are the plane waves: 
\begin{equation}
\vec{E}\left(t,\vec{r} \right) = E_0 \cdot e^{i\vec{k}\vec{r} - i\omega t}  \;\;\; \text{and} \;\;\;   \vec{B}\left(t,\vec{r} \right) = B_0 \cdot e^{i\vec{k}\vec{r} - i\omega t} .
\end{equation}
\subsection{Acceleration without dielectrics}
Interaction between a single plane wave and a charged particle only affects the particles momentum during the interaction. After the interaction, averaging over the electromagnetic field, the particles momentum is unchanged. This is due to the conservation of energy and momentum, sometimes referred to as the Lawson-Woodward theorem for the seven cases Lawson and Woodward postulated under which, if all are satisfied, no net acceleration can take place. \\
One of the seven rules states that nonlinear forces must be neglected. By utilizing nonlinear forces, acceleration without any structures becomes possible. One such force is the ponderomotive force
\begin{equation}
\vec{F}_p = - \frac{e^2}{4m\omega^2}\nabla E^2 .
\end{equation}
It is dependent on the square of the electric field and has the effect of pushing particles from high intensity regions to low intensity regions\cite{Kozak2018a}. By utilizing two or more intersecting laser fields, a travelling interference pattern can be created that is co-propagating with electrons. The properties of this travelling wave, \eg its velocity, are set by the intersecting waves' wavelengths and intersecting angles. By adjusting the mode velocity to the electron velocity, synchronous interaction between the light wave and the electrons can be achieved. One definite advantage of this approach is that due to the absence of any medium, the peak intensity used can be easily increased without damaging structures, however scalability of the interaction distance is not trivial and due to the nonlinear nature of the interaction large pulse energies are required to generate the large peak fields necessary. 

\begin{figure}[ht]
\begin{center}
\includegraphics[width=12cm]{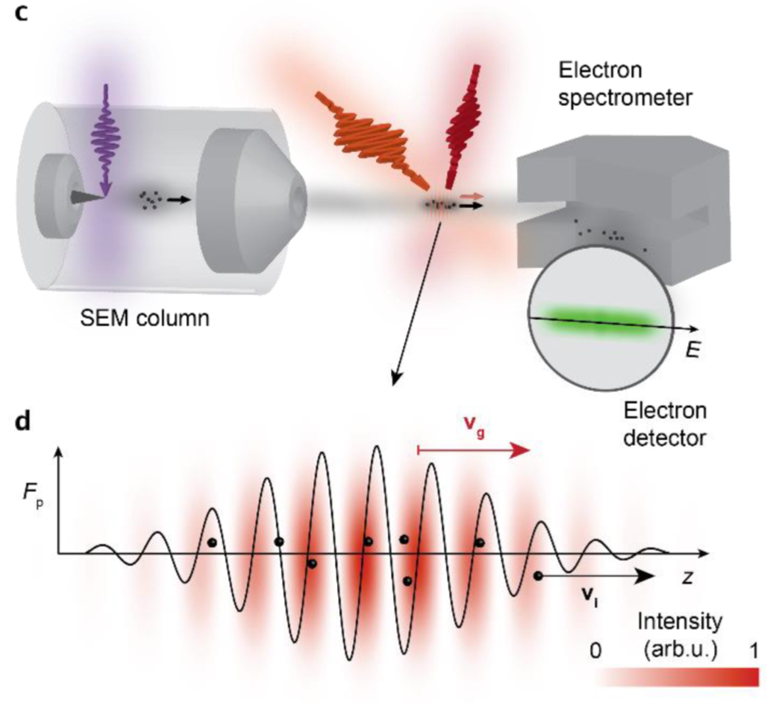}
\caption{Setup to generate a travelling wave to change influence the electrons momenta (top) Ponderomotive wave with group velocity $v_g$ synchronous to the electron velocity. Electrons experience a force according to the gradient from high intensity to low intensity. Electrons on the right hand side of a high intensity region get accelerated, where as electrons on the left side get decelerated. (bottom)}
\label{fig:pond}
\end{center}
\end{figure}

\subsection{Acceleration close to a dielectric interface}
\begin{figure}[ht]
\begin{center}
\includegraphics[width=12cm]{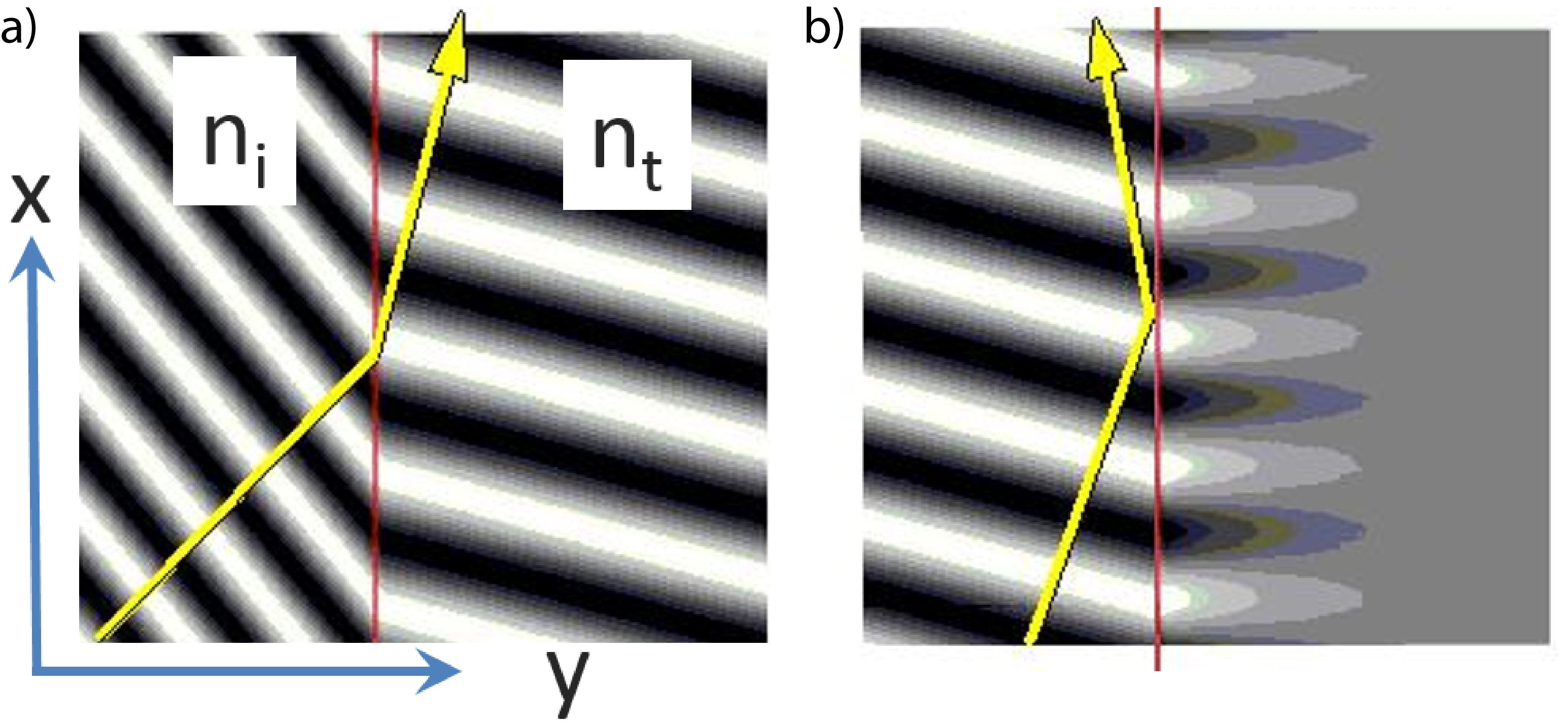}
\caption{Left: refracted wave right: evanescent wave formed by total internal reflection}
\label{fig:waves}
\end{center}
\end{figure}

Placing a dielectric close to the particle beam can alleviate some of the shortcomings of the ponderomotive method. By evaluating the electromagnetic fields in an infinitesimally small area or loop across a boundary, we can arrive at the continuity conditions 
\begin{align}
n_{12} \times \left(\vec{E_2}-\vec{E_1}\right) = 0 \\
\left(\vec{D_2}-\vec{D_1}\right) \cdot n_{12} = \sigma_s\\
\left(\vec{B_2}-\vec{B_1}\right) \cdot n_{12} = 0\\
n_{12} \times \left(\vec{H_2}-\vec{H_1}\right) = \vec{j_s} ,
\end{align}
with $n_{12}$ the normal vector of the interface, $\sigma_s$ the surface charge and $\vec{j_s} the surface current$ -- which both are zero due to only considering dielectrics. From these equations we see that the tangential part of the electric field $\vec{E}$ and the magnetic field strength $\vec{H}$ have to be continuous across the interface. Furthermore, the normal components of the electric displacement field $\vec{D}$ and the magnetic field $\vec{B}$ have to be continuous. \\
Finally, when evaluating these boundary conditions together with the plane waves, we arrive at certain conditions the incident, transmitted and reflected waves have to satisfy. In Fig. \ref{fig:waves} a) we see the incident and transmitted wave at an interface. From 
\begin{equation}
\left( \vec{k_i} - \vec{k_r} \right) \cdot \vec{r} = 0 \;\;\; and \;\;\; \left( \vec{k_i} - \vec{k_t} \right) \cdot \vec{r} = 0
\end{equation}
we see that the x component of the wave vector, parallel to the materials interface, needs to have the same value in all three waves $k_{i,x} = k_{r,x} = k_{t,x}$. We use
\begin{equation}
k_{i,x} = \lvert \vec{k_i} \rvert\sin{\phi} = \frac{n_i\omega}{c} \sin{\phi} ,
\end{equation}
with the dispersion relation $k = \frac{n\omega}{c}$ and the incident angle $\phi$. Similarly we decompose the transmitted wave:
\begin{equation}
\lvert \vec{k_t} \rvert = \frac{n_t \omega}{c} = \sqrt{k^2_{t,x} + k^2_{t,y}}
\end{equation}
Finally we use that $k^2_{t,x} = k^2_{i,x}$ to solve for $k_{t,y}$
\begin{align}
k^2_{t,y} = \left(\frac{n_t\omega}{c}\right)^2 - \left(\frac{n_i\omega}{c}\right)^2 \sin^2\phi .
\end{align}
In the special case that $\phi = \sin^{-1}\frac{n_t}{n_i}$, we get a purely imaginary $k_{t,y}$.
\begin{equation}
k_{t,y} = \pm ik_t \sqrt{\frac{n^2_i}{n_t^2} \sin^2\phi -1} = \pm i\beta k_t
\end{equation}
One of the solutions is non physical and would lead to an exponentially growing field in the y direction. From the other we arrive at the transmitted evanescent wave
\begin{equation}
\vec{E_t} = E_0 e^{-\beta k_t }e^{ik_tx-i \omega t} .
\end{equation}
To use these fields, it is necessary to phase match the electron velocity $v_e = c\beta$ and the phase velocity of the excited evanescent mode $v_{ph} = \frac{c}{n \sin\phi}$. As can be seen this is adjustable via the angle of incidence $\phi$. While possible\cite{Kozak2017}, realizing an accelerator with this technique, starting from subrelativistic electrons, would be challenging. Furthermore there are no other tuning parameters other other than the angle of incidence and the refractive index of the material making more complex field shapes to perform advanced manipulations in phase space virtually impossible. 

Furthermore as can be seen in Fig. \ref{fig:waves} b) the transmitted wave is evanescent, meaning that the y direction decays exponentially. We can evaluate the characteristic decay length to 
\begin{equation}
\Gamma = \frac{c}{\omega\sqrt{n^2 \sin^2 \phi -1}} = \frac{1}{2\pi}\gamma\beta\lambda
\end{equation}
with $\beta$ the electron velocity, $\gamma$ the Lorentz factor and $\lambda$ the wavelength of the light. We see that the spatial extent of the evanescent wave is rather small and therefore the desired high field strengths are only accessible very close to the surface of the dielectric. Therefore the electrons need to occupy a very narrow region above the interface, requiring also a beam of very high quality.\\
To overcome the issue of too few tuning parameters, the interface can be patterned in a periodic fashion. 

\subsection{Acceleration near periodic dielectric structures}
\begin{figure}[ht]
\begin{center}
\includegraphics[width=16cm]{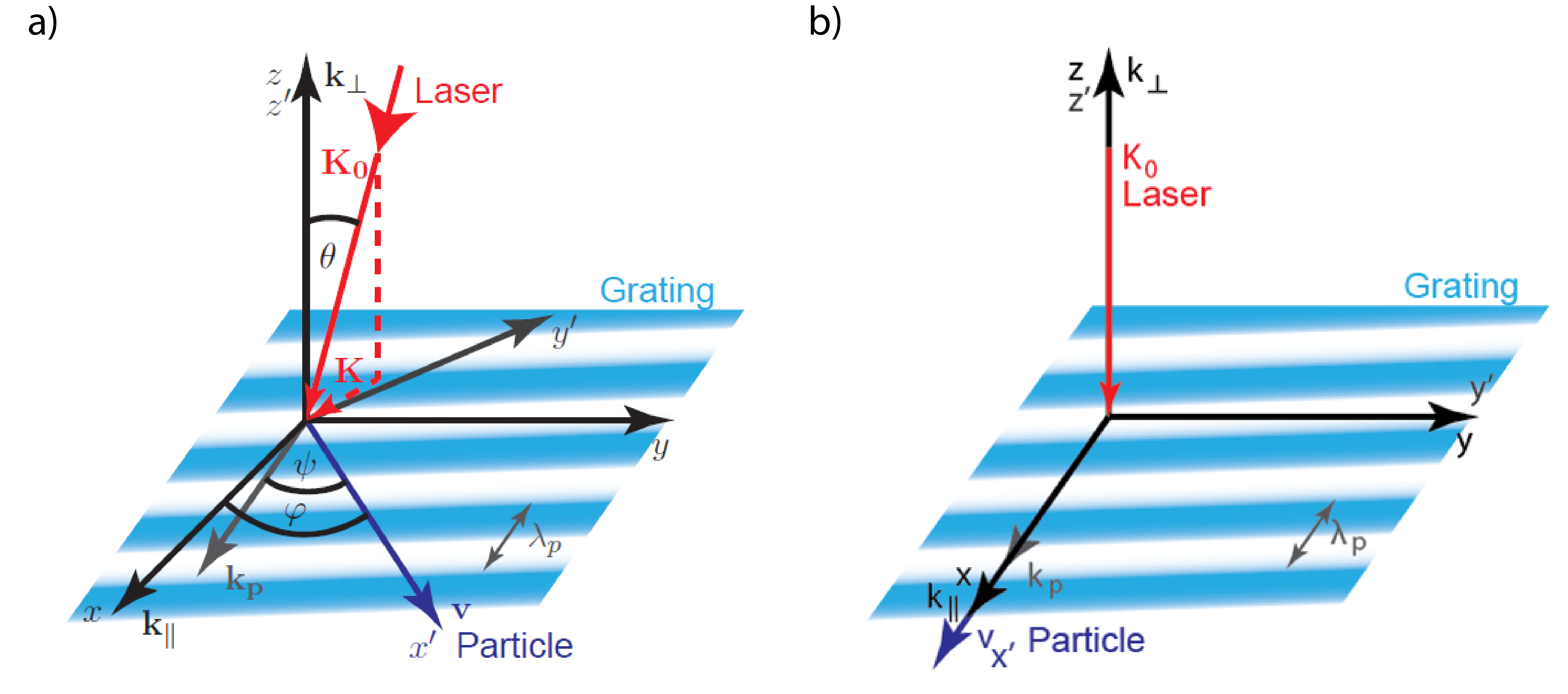}
\caption{Left: general case of light illuminating a periodic grating Right: simplified case Taken from~\cite{Breuer2014}}
\label{fig:grating}
\end{center}
\end{figure}

When considering the general case of a laser illuminating an infinite plane grating of periodicity $\lambda_p$ many parameters have to be considered, as seen in Fig. \ref{fig:grating} a). By aligning all the coordinate systems and choosing the laser incident angle to be perpendicular to the grating we can drastically simplify the arrangement. Here $\vec{K_0}$ is the incident wave vector, $\vec{K}$ the component parallel to the surface and $\vec{k_\parallel}$ and $\vec{k_\bot}$ the parallel and perpendicular diffracted components respectively. The most interesting part for the study of acceleration is the parallel diffracted component  $\vec{k_\parallel}$. This can have multiple spatial harmonics  $\vec{k_\parallel^n} = \vec{K} + n\vec{k_p}$.

The grating fields can be described as a series of these spatial harmonics.
\begin{equation}
\vec{A}(\vec{r},t) = \sum^{\infty}_{n=-\infty} \vec{A_n} e^{i(k_\bot^nz+k_\parallel^n r-\omega t+\phi)}
\end{equation}

To achieve phase matching, the same condition must be satisfied as for the case of evanescent fields at a plane dielectric interface. The electron velocity and the mode velocity of the diffracted light wave must be the same. Here the mode velocity is given by $v_{ph} = \omega/k_parallel \cos\phi$ resulting in 
\begin{equation}
k_\parallel = \frac{\omega}{\beta c \cos\phi} = \frac{k_0}{\beta \cos \phi}
\end{equation}
Assuming that the particle trajectory is parallel to the grating vector, we arrive at the synchronicity condition 
\begin{equation}
\lambda_p = n\beta \lambda .
\end{equation}

Using $\vec{k_\parallel}$ and $\vec{k_\bot}$ in Ampere's and Faraday's law, we obtain
\begin{equation}
\vec{E} = \left(
\begin{array}{c}
icB_y/(\tilde{\beta}\tilde{\gamma})\\
E_y\\
-cB_y/\tilde{\beta}\\
\end{array}
\right)
\end{equation}

and 

\begin{equation}
\vec{B} = \left(
\begin{array}{c}
icE_y/(\tilde{\beta}\tilde{\gamma})\\
B_y\\
E_y/(\tilde{\beta}\tilde{\gamma})\\
\end{array}
\right) .
\end{equation}

With the commonly known equation for the Lorentz force $\vec{F} = q(\vec{E} + \vec{v} \times \vec{B})$ we arrive at the resulting force from these electromagnetic fields
\begin{equation}
\vec{F} = \left(
\begin{array}{c}
icB_y/(\beta \gamma)\\
0\\
-cB_y/(\beta \gamma^2)\\
\end{array}
\right) .
\end{equation}

\begin{figure}[ht]
\begin{center}
\includegraphics[width=16cm]{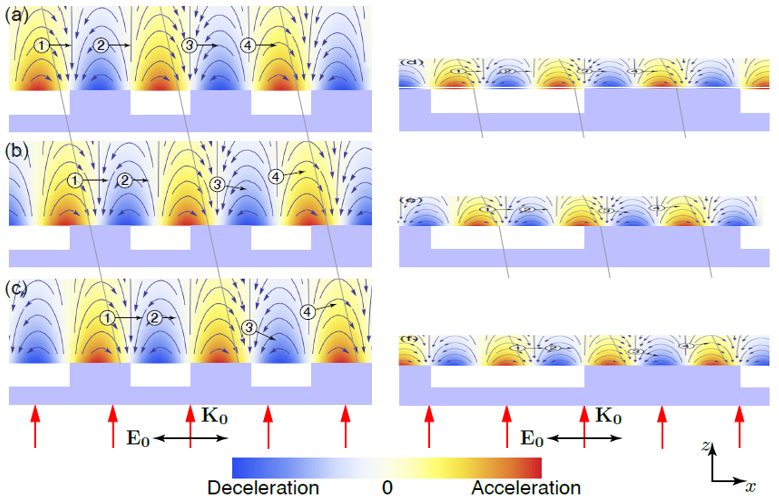}
\caption{Left: First spatial harmonic excited at a dielectric grating. Right: Third spatial harmonic excited at the same grating. The x coordinate is stretched for easier visualization. }
\label{fig:fields}
\end{center}
\end{figure}

The resulting forces are shown in Fig. \ref{fig:fields}. It is immediately apparent that there is an invariant direction, where no forces are present. This direction is parallel to the grating teeth.\\
Furthermore the mode still decays exponentially in the z direction. However the phase matching is now accomplished via the periodicity $\lambda_p$ of the grating. This exponential decay has the side effect that particles that get pushed away from the grating surface due to the inherent transversal forces, can never be recaptured by the transversal forces, even if the phase is flipped and the force points towards the grating, since the force at a greater distance is much weaker.\\
Also like in the case of the evanescent wave at the interface, the peak field needed to accelerate electrons is much lower than when using nonlinear effects like the ponderomotive force. This is beneficial since any structure can be damaged. 

\begin{figure}[ht]
\begin{center}
\includegraphics[width=16cm]{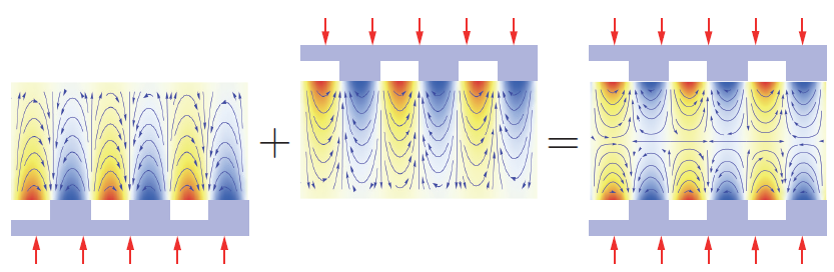}
\caption{Combining two opposing gratings and their fields yields a configuration where the field is now also symmetric in the y direction, removing the complications of the evanescent decay of the fields.}
\label{fig:dual}
\end{center}
\end{figure}

The other major drawback of this type of structure is that apart from the evanescent decay, no speed of light mode is supported. That means that it is not possible to accelerate relativistic electrons at a single sided structure such as a dielectric interface or a single dielectric grating. Both these problems can be solved when combining two gratings. The result can be seen in Fig. \ref{fig:dual}. There two gratings are combined, opposing each other. The overlapping evanescent fields form a hyperbolic cosine or sine shape, depending on the phase of both incident lasers. While the usable spatial extent of the fields is not much increased, the gradient of the field decay is much less. Therefore electrons occupying the space between the gratings experience a much more uniform energy modulation as in the same electron population would fly by a single grating.\\
Furthermore the double sided grating supports a speed of light mode, enabling the acceleration of relativistic electrons and also deflecting modes, when the incident lasers' phases are adjusted. More information is available in \cite{Breuer2014}.

\pagebreak

\subsection{Guiding forces for extended interaction between particles and laser fields}
\begin{figure}[ht]
\begin{center}
\includegraphics[width=16cm]{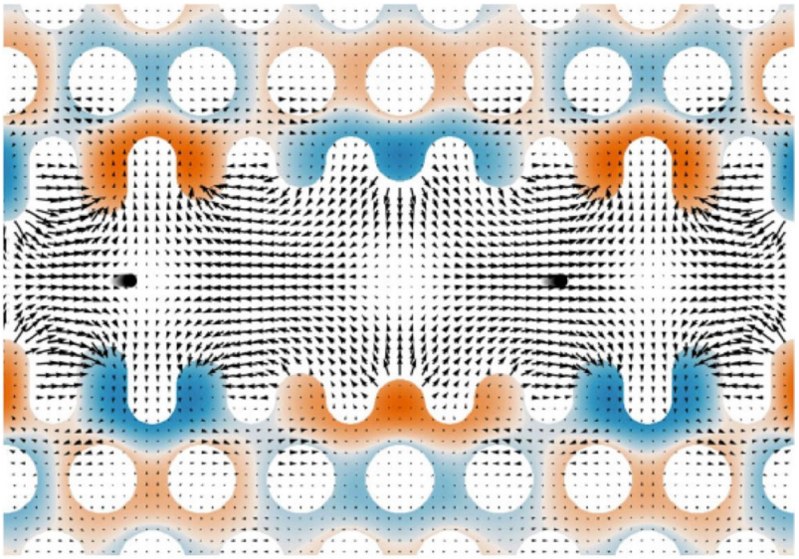}
\caption{Galaxie structure: Not included in the picture are waveguides located at the top and bottom of the here visible region to supply the optical power. The hole pattern is part of a photonic crystal used as the waveguide. Similar to the previously shown simple dual gratings, two opposing gratings generate a non evanescent field profile between them. In this case two periodicities are visible. The ``tooth to tooth'' periodicity exiting the mode that accelerates the electrons and a macro periodicity responsible for exciting the mode that confines the electrons. The shown fields are the superposition of both the accelerating and the confining mode. More information and animations of the fields and phase space can be found at \cite{naranjoweb}}
\label{fig:galaxie}
\end{center}
\end{figure}

As briefly mentioned in the previous chapter, the transversal forces present in any DLA can deflect the particles towards or away from the centre of the grating surfaces. Even in the case of the double sided grating, where the fields are symmetric, there is only one point in parameter space, where no transversal force is present. This is perfectly on crest. Apart from it being virtually impossible to inject electrons only perfectly on crest and keeping them there during the acceleration, manufacturing tolerances, laser amplitude fluctuations and other perturbations would destroy this balance.\\
Therefore another mechanism is required to make sure electrons are not deflected so much that they crash into the structures and cause dramatic beam loss. There are different mechanism available to achieve this goal. \\
One such technique is explored numerically by B. Naranjo\cite{Naranjo2012} as part of the Galaxie proposal for a photonic accelerator. Here multiple modes are excited, similar to the spatial modes mentioned previously. One mode, the synchronous one, is used for acceleration. However the modulation in the depth of the grating teeth seen in Fig. \ref{fig:galaxie} causes the emergence of a second mode. This mode is non synchronous to the electrons however when averaged over longer distances a net force appears that creates a moving potential minimum and thus confines the electrons.

\begin{figure}[ht]
\begin{center}
\includegraphics[width=12cm]{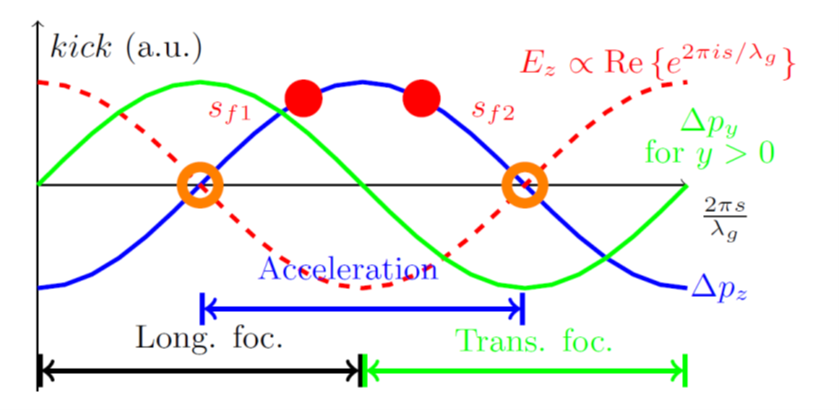}
\caption{The shown graph illustrates the APF mechanism. The blue curve represents the longitudinal kick an electron receives per unit cell of the periodic accelerator structure. The green curve shows the transverse kick. In theory if it were possible to hold the electrons exactly on crest of the accelerating field, no transverse forces appear. However small manufacturing defects or perturbations will cause enough phase differences so this is not achievable. Choosing two design points \ie the red points only slightly off crest, one looses only little acceleration force but periodically varying between the two points gives on average a force that keeps the particles confined in the channel. Electrons can also be just guided without experiencing acceleration by choosing the orange circles as design points. \cite{Niedermayer2018}}
\label{fig:apf}
\end{center}
\end{figure}

Another method, currently being evaluated experimentally as of the writing of these proceedings, is an adaptation of alternating phase focusing. This method was previously used for RF accelerators. Compared to the previous method where two separate modes need to be excited, alternating phase focusing (APF) makes use of the inherent deflecting forces present in the accelerating mode. By choosing two operating points that are slightly off rest in terms of acceleration, depicted in Fig. \ref{fig:apf} by the red circles, and periodically toggling between these points, the design electrons experience both transversally focusing and defocusing forces, which in total keep the electrons confined. For more information see \cite{Niedermayer2018}. 

\begin{figure}[ht]
\begin{center}
\includegraphics[width=16cm,trim={2cm 12cm 3cm 11cm},clip]{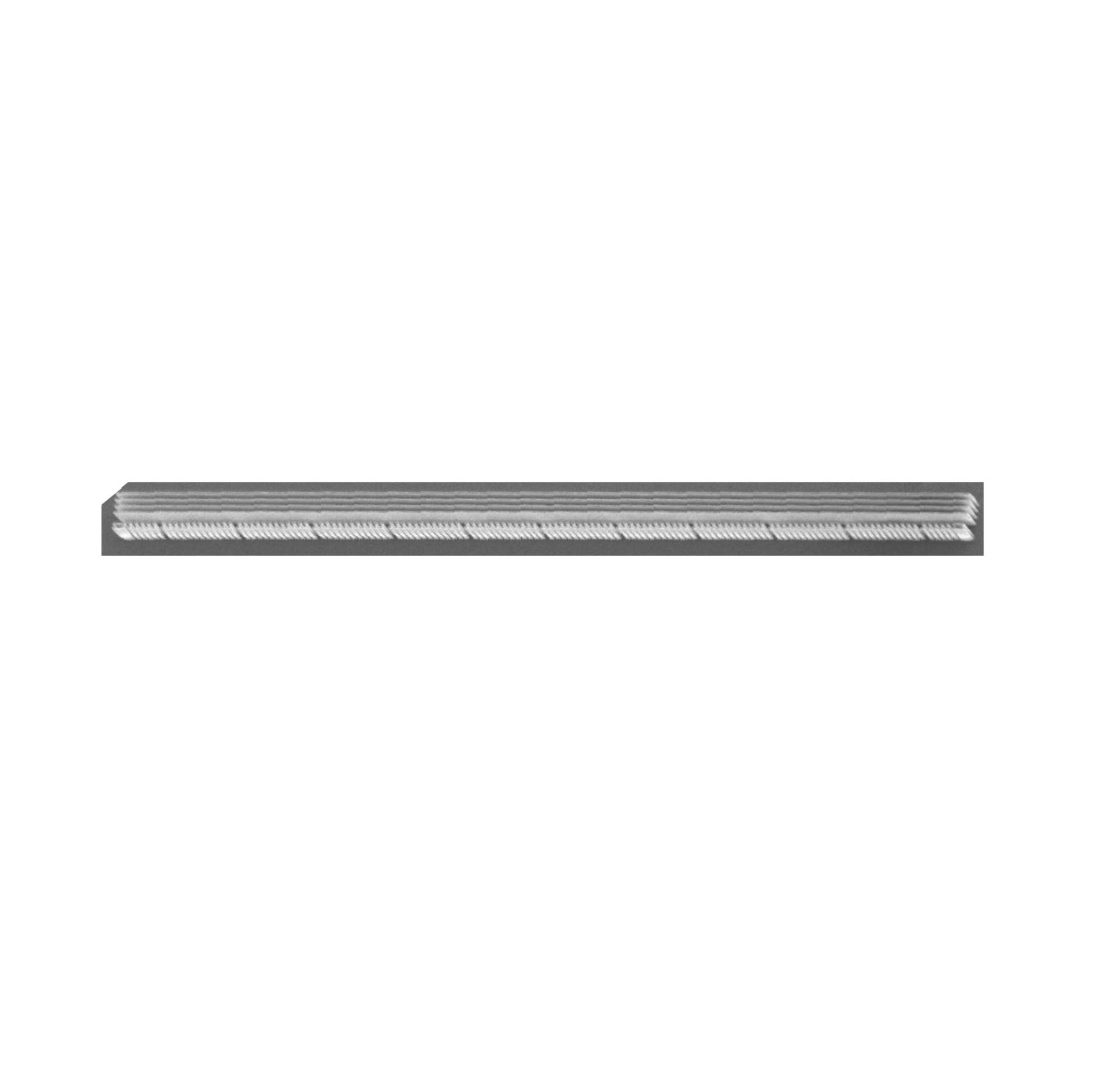}
\caption{Implemented APF design. Visible are a Bragg reflector at the top of the structure. The 4 dielectric blades act as a mirror and reflect the laser light, incident from the bottom, back. This mimics double sided illumination. Furthermore the periodicity of the accelerating structure and the periodic phase jumps are visible. This \Unit{80}{$\mu$m} long structure is the longest structure tested at sub relativistic speeds to date.}
\label{fig:apfstruct}
\end{center}
\end{figure}

A Structure applying this principle can be seen in Fig. \ref{fig:apfstruct}. Here periodic notches in the accelerating structure are used to introduce the required phase jumps. Preliminary tests show good agreement between theory and experiment proofing the viability of this method. With appropriate structures even the until now invariant direction can be controlled.

\begin{figure}[ht]
\begin{center}
\includegraphics[width=16cm]{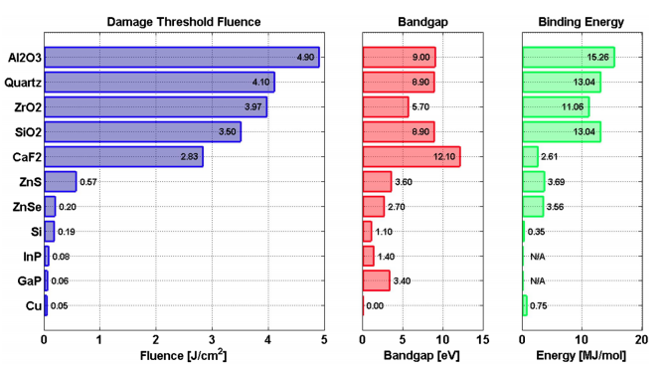}
\caption{Overview of material parameters. Taken from~\cite{Soong2012}}
\label{fig:table}
\end{center}
\end{figure}

Lastly an important choice is the material of the structures. It must fulfil several criteria. First of all it must be manufacturable at the sizes required for a DLA. Some processes are very well developed in terms of structure fabrication, mostly due to the semiconductor industry. However other materials like Al$_2$O$_3$ or Diamond can be very hard to manufacture. \\
Secondly, the chosen material should have a damage threshold as high as possible. A selection of materials can be seen in Fig. \ref{fig:table}. For example while silicon is relatively easy to manufacture its damage threshold for laser induced damage is fairly low. If suitable processes are available it could be beneficial to use different materials.\\
However the damage threshold does not paint a complete picture. Especially at subrelativistic speeds, the mode excitation efficiency, so ho much of the incident laser light is diffracted into the usable accelerating mode is dependent on the refractive index. The higher the refractive index the better the mode excitation efficiency. While it might be beneficial for damage threshold reasons to choose a different material than \eg silicon, it might be that even though a higher incident laser peak field can be used, less field is coupled to the relevant mode, hurting the efficiency. Therefore a delicate balance must be struck between all material parameters.

\section{DLA Experiments and state of the art technology}
\subsection{Acceleration and electron control}
\begin{figure}[ht]
\begin{center}
\includegraphics[width=16cm]{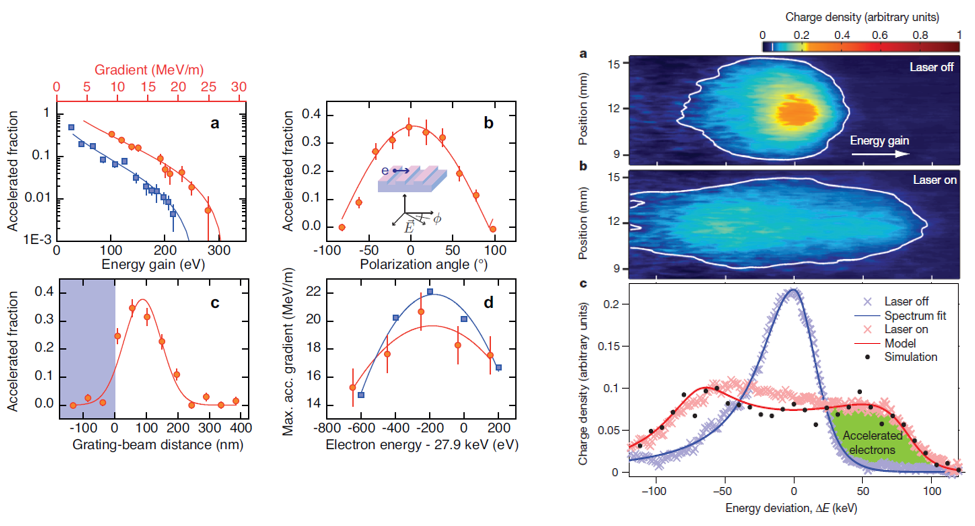}
\caption{First results of dielectric laser acceleration both at subrelativistic electron velocities (left) and relativistic electron speeds (right). The left results show the spectrum of the energy modulated electrons and dependence on the polarization angle, grating-beam distance and initial beam energy. In the cases of the polarization angle the available x component of the field is diminished when the polarization is changed from the optimum and hence electrons are accelerated less. The same holds when the beam-grating distance is increased. When decreasing the distance, the beam is eventually clipped on the structure. Changing the initial electron energy changes the phase matching and hence the maximum acceleration gradient. The relativistic results show a full modulation of the complete electron population. Furthermore the modulates spectrum shows the onset of the characteristic double horn spectrum. This is an indicator that the fields are of good quality. Taken from~ \cite{Breuer, Peralta2013}}
\label{fig:proof}
\end{center}
\end{figure}

As mentioned in the introduction, the initial experiments at optical wavelengths were performed in 2013. One Experiment was executed at \Unit{30}{keV} electron energy in a modified electron microscope\cite{Breuer}. Here a single sided fused silica grating was used in conjunction with a Ti:Sapphire laser. Due to the subrelativistic nature of the electrons and the laser wavelength using the synchronicity condition Eq.~(24) the required periodicity would have been around $\lambda_p = 260nm$ which was not available at the time. Therefore, the third spatial harmonic was chosen to drive the interaction. Although possible, the excitation efficiency of the higher order spatial modes is very low. Also the relatively low refractive index of fused silica affects the mode excitation efficiency negatively. Therefore the achieved acceleration gradient was limited to \Unit{25}{MeV/m}. \cite{Breuer}.\\
The demonstration at relativistic energies --- performed at SLAC's NLCTA\cite{Peralta2013} --- did not suffer these shortcomings as the increased electron velocity at \Unit{60}{MeV} was already relativistic, relaxing the structure parameters by a factor of three. However since $\beta = 1$ in this case, a double sided grating needed to be used, to generate a speed of light mode. With this, acceleration gradients of more than \Unit{250}{MeV/m} were achieved. Some results are shown in Fig. \ref{fig:proof}.

\pagebreak

\begin{figure}[ht]
\begin{center}
\includegraphics[width=16cm]{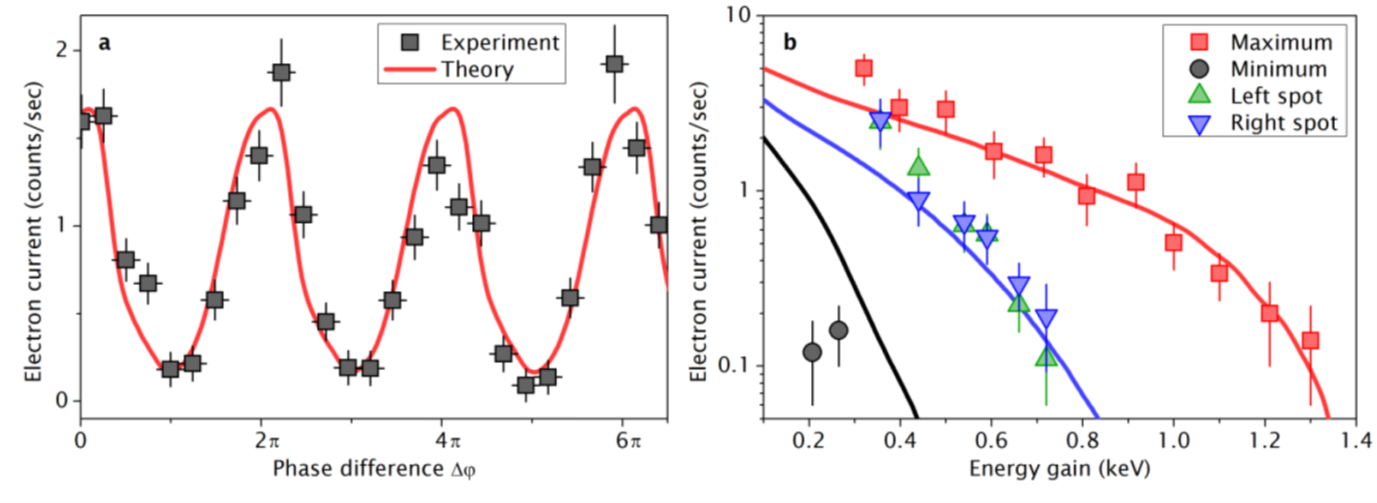}
\caption{Proof of principle experiment for the staging of multiple DLAs. Left: Phase dependence of the final energy. Right: Measurement of each individual interaction (blue and green) at a phase difference of the two laser pulses where destructive interference is reached (black) and similarly at a phase difference where there is constructive interference, doubling the final energy of the electrons. Taken from~cite{McNeur2018}}
\label{fig:staging}
\end{center}
\end{figure}

After these initial demonstrations it was concluded that for the subrelativistic side, structures made of silicon in conjunction with near infrared (NIR) laser sources in the regime of $\lambda = $\Unit{2000}{nm} are most suitable. With these, many important steps could be shown.\\
Staging is arguably one of the most important properties of any accelerator. It needs to be possible to concatenate accelerators without loosing performance. It was shown at a single sided dielectric grating that staging without performance loss is possible \cite{McNeur2018}. Here two identical laser illumination spots were placed on a grating. As seen in Fig.~\ref{fig:staging} in the blue and green curve, each interaction on their own causes the same energy modulation. When both stages are illuminated simultaneous, it is dependent on the phase of the two laser spots what final energy of the electrons is reached. When the two pulses are completely out of phase, the energy gain contributed by the first interaction is almost completely negated by the second. Analogous when the two interactions happen in phase, the energy gain is almost doubled. The small remaining energy and not quite energy doubling is attributed to the transversal forces. Electrons that experience a transversal kick away from the grating surface in the second interaction will experience a weaker force in the second, even if the phase is spot on. Due to the exponential decay of the fields there is a big gradient even for small distance changes towards or away from the grating, as shown in Fig.~\ref{fig:proof}.

Another necessary requirement of an accelerator is steering of the particle beam. As we've already seen, deflecting forces are easily available in the transverse direction perpendicular to the grating surface. Furthermore no forces are present in the other transverse direction parallel to the grating. This can simply be remedied by not assuming the most simplified case as we did in the theory part but allow for an angle between the particle propagation direction and the grating vector to be present, \eg $\psi \neq 0$ in Fig.~\ref{fig:grating}. This leads to a different force the particle feels in its reference frame:
\begin{equation}
\vec{F} = q\left(
\begin{array}{c}
icB_y/(\tilde{\beta} \tilde{\gamma}) + \tan\phi E_y\\
-icB_y \tan\phi /(\tilde{\beta} \tilde{\gamma}) - \tan\phi \sin\phi E_y\\
-cB_y (1-\tilde{\beta}^2)/(\tilde{\beta} ) + i\tan\phi E_y/\tilde{\gamma}\\
\end{array}
\right) .
\end{equation}
The force in y direction is now non zero. This can be used to deflect electrons \cite{Kozakdefl}. A deflection of up to \Unit{6}{mrad} was achieved over an interaction distance of roughly \Unit{10}{$\mu$m}. The same principle can be applied to construct a lens. Using parabolic grating teeth a position dependent grating angle can be introduced acting as a deflector of different strength depending on the entrance offset of the beam with respect to the symmetry axis of the parabolic grating\cite{McNeur2018}. Since the input beam size was very small with respect to radius of curvature of the grating teeth, each interaction of one beam position with the grating could be viewed as a pure deflection. A knife edge is used behind the structure to determine the amount of deflection. This raytracing approach allowed for the measurement of the focal length of the structure. Obviously this lens only works for a certain energy range and is fairly chromatic.

\begin{figure}[ht]
\begin{center}
\includegraphics[width=14cm]{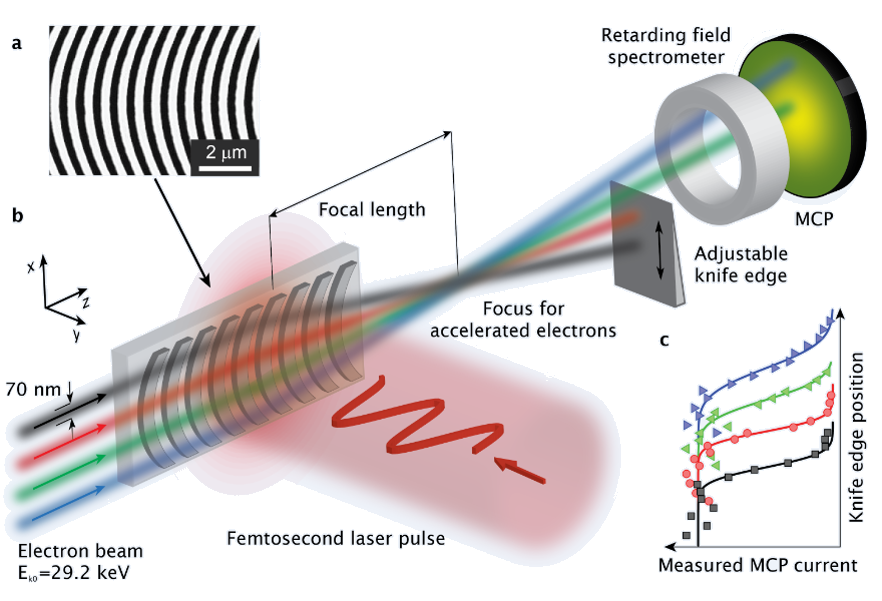}
\caption{Lensing structure and corresponding data. Taken from~\cite{McNeur2018}}
\label{fig:lens}
\end{center}
\end{figure}

\begin{figure}[ht]
\begin{center}
\includegraphics[width=16cm,trim={2cm 0cm 0cm 0cm},clip]{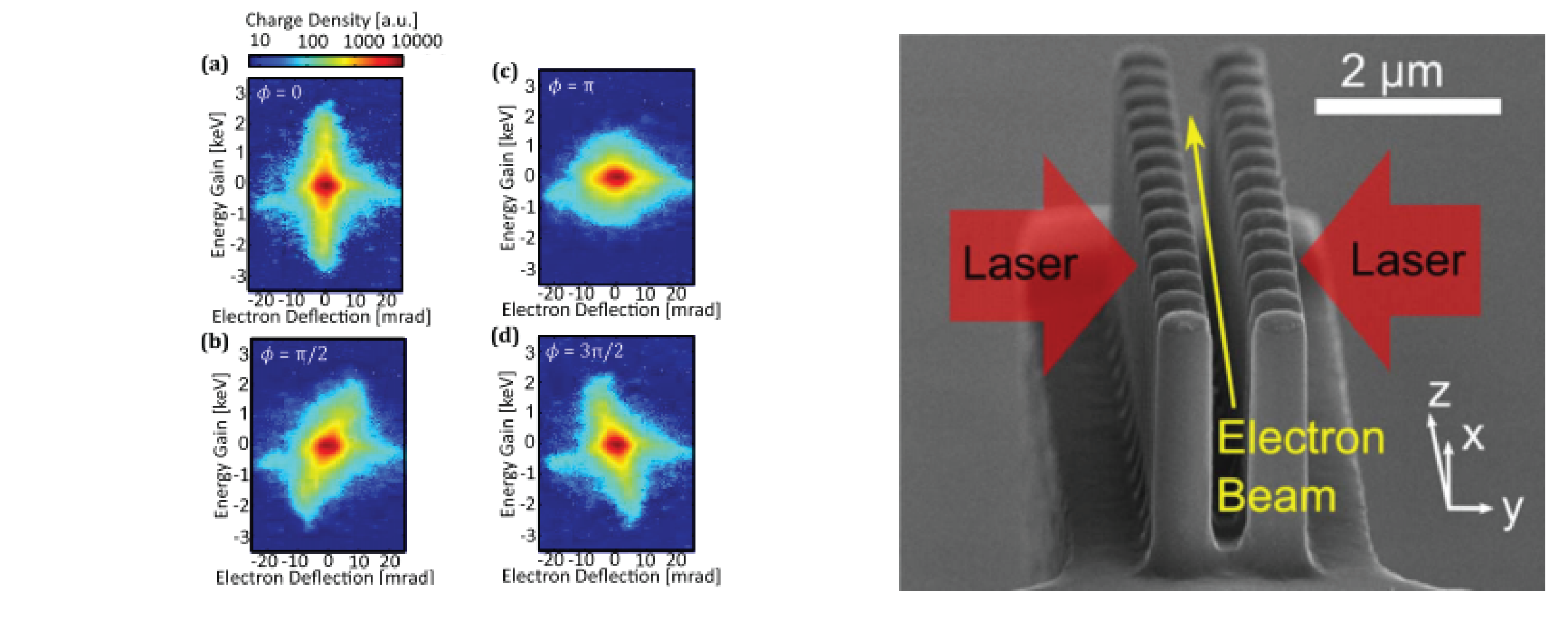}
\caption{Phase dependant deflection of electrons (left) and dual pillar structure geometry (right). Taken from~\cite{leedle2018}}
\label{fig:leedle2018}
\end{center}
\end{figure}

As alluded to in the theoretical description, the phase difference of two lasers driving a double sided grating can also affect the mode that is present in the device. The modes can be switched between an accelerating and a deflecting mode\cite{leedle2018}. Here a new type of structure was used, seen on the right of Fig.~\ref{fig:leedle2018}. To ease the manufacturability of double sided gratings, the original approach of etching two separate grating and aligning them afterwards, which is quite demanding and prone to inaccuracies, was exchanged for structures that can be produced in one etching step, with the accuracy limited by the tolerances of the manufacturing machines. These structures consist of two rows of silicon pillars. If etched deep enough they approximate an infinitely wide double sided grating. In the shown data in Fig.~\ref{fig:leedle2018} one can observe the dependency of energy gain and maximum deflection angle on the relative phase of the electrons. \\
In the case of $\Psi = 0$ the acceleration is maximised with minimal deflection. Opposite for $\Psi = \pi$, the deflection is maximised. Phase difference values in between lead to a mixed state of acceleration and deflection.

The current record gradients achieved by dielectric laser accelerators are \Unit{210}{MeV/m} at $\beta = 0.3$ \cite{Kozak2017b}, \Unit{370}{MeV/m} at $\beta = 0.7$\cite{Leedle2015} and \Unit{850}{MeV/m} with \Unit{6}{MeV} electrons\cite{Cesar2018}. The current limitation for nonrelativistic electrons are the poor damage threshold of silicon structures and the mode excitation efficiency. Different  material and geometry choices can improve the gradient considerably.

\begin{figure}[ht]
\begin{center}
\includegraphics[width=16cm]{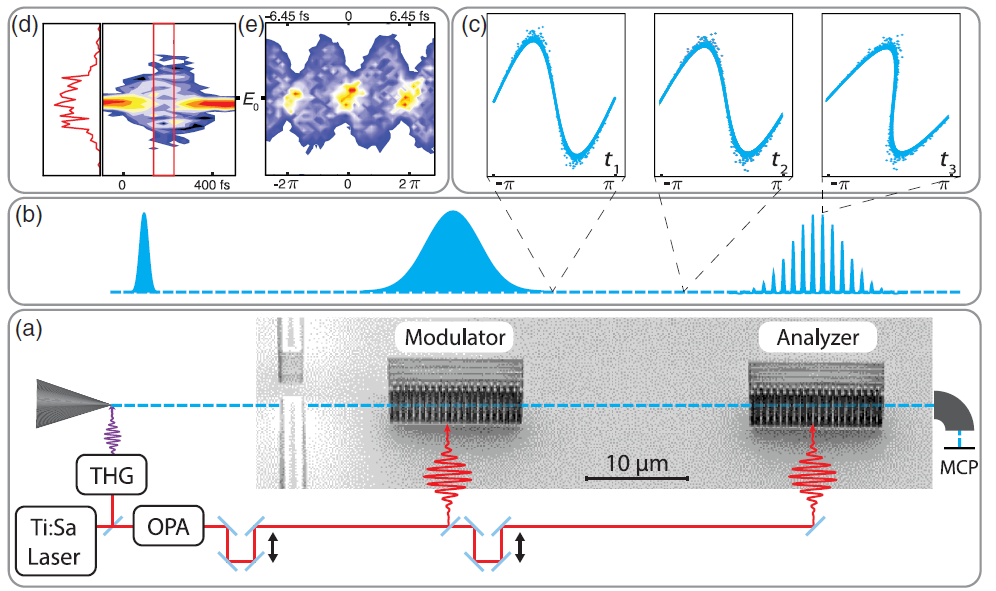}
\caption{(a) Setup for the generation and characterization of attosecond electron micro bunch trains. In the first interaction region, electrons are imprinted with a sinusoidal energy modulation that during a drift forms a density modulation in the longitudinal direction. The second interaction is used to characterize the micro bunch trains. (b) Longitudinal electron density over the whole length of the experiment. 100 fs long electron pulses are generated via photoemission at the electron emitter. During propagation various effects broaden the pulse to roughly 400 fs. After the first interaction a micro bunch train is formed. (c) Evolution in phase space from the initial sinusoidal energy modulation to the ideally bunched beam. Fast electrons catch up with slow electrons to form the micro bunches. (d) Electron spectrum after a single interaction vs relative delay between electron pulse and laser pulse. (e) Elektron spectrum after the analyzer vs relative phase of the two laser pulses. This spectrum can be compared to simulations to reconstruct the micro bunch length. Taken from~\cite{Schoenenberger2019}}
\label{fig:bunching}
\end{center}
\end{figure}

All experiments at subrelativistic electron speeds shown thus far were performed over short distances of a few micron. As discussed in the theoretical section on beam guiding there are known mechanisms of how to keep the electrons inside the accelerator channel. However for these mechanisms to work, the electrons need to inhabit a small region in phase space. This can be achieved in the structures themselves or externally through micro bunching. \\
Micro bunching happens in subrelativistic electrons when a sinusoidal energy modulation is imprinted on a set of particles. The electrons with higher energy have a higher velocity. Analogous electrons with reduces energy will be slower. This is depicted in Fig.~\ref{fig:bunching}. While drifting through vacuum, the faster electrons catch up with the slower ones, creating a higher electron density in real space. This mechanism is also often called ballistic microbunching, since no fields are applied after the initial modulation and the electrons propagate ballistically. \\
With this scheme, electron micro bunch durations of \Unit{700}{as}\cite{Black2019} and \Unit{270}{as}\cite{Schoenenberger2019} have been shown. These bunches are short enough to be injected into another accelerator.

\subsection{Laser power delivery}

As mentioned above, all the shown experiments and even all experiments concluded, that the authors are aware of as of the time of writing, have been performed over distances of at most \Unit{1}{mm}. This is still possible via a single elliptical laser spot. When building longer devices, using free space laser coupling to the structures is no longer feasible. 

\begin{figure}[ht]
\begin{center}
\includegraphics[width=15cm]{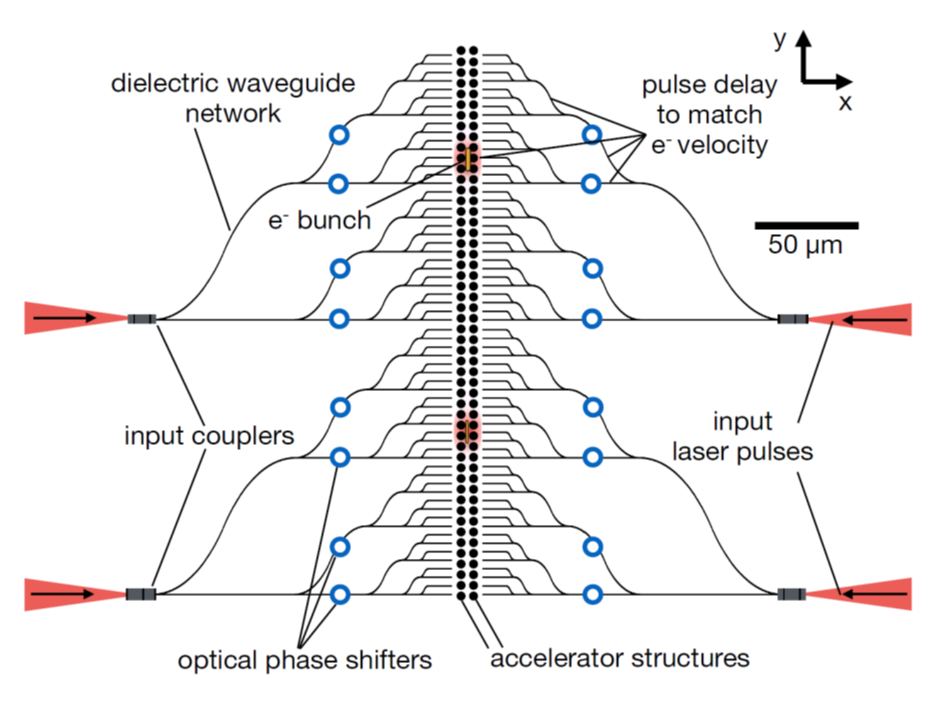}
\caption{Laser power delivery network. Laser light is coupled to a waveguide structure via input couplers. The encoded length of the branch structure synchronizes the laser arrival to the electrons. Optical phase shifters can be used to fine adjust the optical phase. Taken from~cite{Hughes2018}}
\label{fig:hughes}
\end{center}
\end{figure}

Therefore another scheme to distribute power to the accelerator is needed. Similar to RF waveguides in RF accelerators, photonic waveguides can be used to distribute the power.  Case studies of how such a system might be designed, which tuning knobs are available and where the limitations lie have been done. The result of one such study is shown in Fig.~\ref{fig:hughes}. The accelerating structure is fed by a tree branch structure of dielectric waveguides. The necessary delay between the electrons and the incoming laser is encoded into the structure and can be finely controlled via included optical phase shifters to adjust for fabrication inaccuracies or thermal expansion.

These structures still require many inputs for the delivered power since the waveguides, constructed form silicon or a derivative material such as silicon nitride, have a fairly low damage threshold. Adopting technologies such as photonic crystal waveguides and on chip pulse compression might alleviate the stress on the waveguides. Furthermore, the currently used wavelength of \Unit{2000}{nm} was not chosen at random. Fiber lasers are available at these wavelengths that can still produce short pulses. Hence a network of fiber lasers can be used to drive multiple inputs on these structures. 

Other research was focussed on the individual components such as the beam splitters at every tree branch. The shown example is a wavelength demultiplexer that takes an input of multiple wavelengths and splits them into separate channels. This is easily adapted to splitting power instead of wavelengths. These devices are attractive for their small size. The very unintuitive design is generated by so called inverse design. Here a design boundary and the inputs as well as the desired output are supplied. The inverse design algorithm adjusts the geometry in the design region to achieve the desired goal.

\begin{figure}[ht]
\begin{center}
\includegraphics[width=16cm]{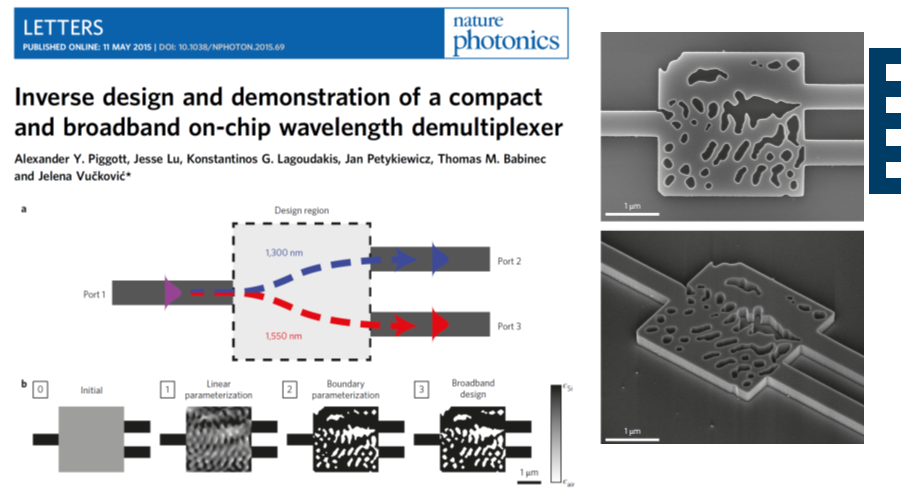}
\caption{Photonic structure designed with the help of inverse design. Taken from~\cite{Piggott2015}.}
\label{fig:invdes}
\end{center}
\end{figure}

Finally a recycling of the laser power can be envisioned. Most of the laser pulse leaves the structure without giving energy to the electrons. Therefore the efficiency is quite poor. If it were possible to include the accelerator in a resonator cavity including a gain material, the optical power could be recirculated\cite{Siemann2004,NeilNa2005}.

\subsection{Electron beam requirements}
Lastly we discuss the requirements the DLAs place upon the electron beams. From estimates conducted in \cite{Breuer2014} and \cite{Niedermayer2018}, DLAs require normalized emittances of sub nm-rad. These emittances are usually only approached by electron microscopes, where excellent beam quality is needed to achieve the best resolutions. This however is achieved by filtering the electron beam until the quality is good enough. This means that such a source is not necessarily well suited for the application in DLAs. Since the sources are laser triggered, starting with a high emittance beam and filtering can cause pulse prolongation due to space charge in the gun. An optimized emitter that emitts less electrons at a smaller emittance might be advantageous. Research into many different types of emitters is being conducted. This includes low emittance rf photoinjectors\cite{Li2012}, diamond coated silicon pyramids\cite{Simakov2017}, diamond coated tungsten tips\cite{Tafel2019}, LaB$_6$ nano wires\cite{Zhang2016} and silicon tips and tip arrays.

\begin{figure}[ht]
\begin{center}
\includegraphics[width=16cm]{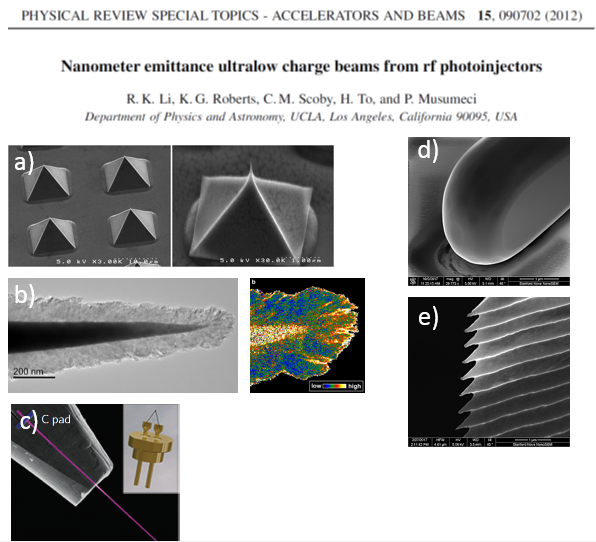}
\caption{Overview over some new electron emitter designs that show promising properties for DLA applications: low emittance rf photoinjectors\cite{Li2012}, diamond coated silicon pyramids\cite{Simakov2017}, diamond coated tungsten tips\cite{Tafel2019}, LaB$_6$ nano wires\cite{Zhang2016} and silicon tips and tip arrays}
\label{fig:sources}
\end{center}
\end{figure}

\section{Conclusions}
DLAs have come far since their proof of principle experiments in 2013. While they will likely not find application as high energy particle physics machines due to the limited current available, they are a unique tool in the world of novel accelerators because they don't rely on huge infrastructure such as existing accelerators or petawatt class lasers. Combined with the small size of the accelerator itself, various new opportunities are opening up that other accelerator technologies might not be suitable for. One example are medical devices. A miniaturized accelerator could be used endoscopically to irradiate tumours directly and reducing damage to surrounding tissues.\\
Table top radiation sources using the principles of free electron lasers, Smith-Purcell radiation or inverse Compton scattering are feasible and could supply smaller research facilities with the means to not be dependent on the admittedly growing but still very limited number of FEL facilities.\\
Lastly the inherent time scales make DLAs well suited for the development of ultrafast machines for electron microscopy and diffraction. As shown through simple methods very short electron bunches can be created. In this form DLAs might also become useful for large scale facilities as a means to modulate electron bunches on tiny time and length scales.

\end{document}